\documentclass[a4paper,12pt]{article}
\usepackage{jheppub_modified_green}
\usepackage{notoccite}
\usepackage{mathtools}
\usepackage{comment}

\newcommand{\lt}{\tilde{\lambda}}
\newcommand{\tLam}{\tilde{\Lambda}}
\newcommand{\lan}{\langle}
\newcommand{\ran}{\rangle}
\newcommand{\vev}[1]{\langle{#1}\rangle}
\newcommand{\bev}[1]{[{#1}]}
\newcommand{\pa}[1]{\Big(#1\Big)}
\newcommand{\sg}{\operatorname{sg}}
\newcommand{\cA}{\mathcal{A}}
\newcommand{\cN}{\mathcal{N}}

\newcommand{\bq}{\bar{q}}
\newcommand{\bsb}{\bar{s}}

\def\be{\begin{equation}}
\def\ee{\end{equation}}

\def\Cdot{{\cdot}}

\allowdisplaybreaks

\title{$\cN=4$ single-minus superamplitudes
and dual superconformal symmetry}

\author{Andreas Brandhuber,}
\author{Paolo Pichini,}
\author{Gabriele Travaglini}
\author{\\and Congkao Wen}
\affiliation{Centre for Theoretical Physics and Astronomy, Department of Physics and Astronomy,\\
Queen Mary University of London,
Mile End Road, London E1 4NS, United Kingdom}
\emailAdd{a.brandhuber@qmul.ac.uk}
\emailAdd{p.pichini@qmul.ac.uk}
\emailAdd{g.travaglini@qmul.ac.uk}
\emailAdd{c.wen@qmul.ac.uk}

\begin{document}
\begin{flushright}
QMUL-PH-26-10
\end{flushright}

\abstract{

We construct the $\cN{=}4$ supersymmetric completion
of the recently proposed single-minus gluon
amplitudes in  $(2,2)$ signature, which 
are nonvanishing for all multiplicities 
on a  half-collinear kinematic locus.
The superamplitude factorises into
a  permutation-invariant  measure 
$\Delta^{(n-1)}$ with uniform little-group weight that imposes the half-collinearity constraint,
a piecewise constant stripped amplitude
$\tilde{A}_{1\ldots n}$ that is helicity blind
and dual conformal invariant,
and  (super)momentum conservation
delta functions.
For $n{=}3$, our superamplitude  reduces to the known
$\overline{\rm MHV}$ superamplitude.
We prove dual superconformal covariance of the $n$-point superamplitude, 
and further analyse the $\mathrm{Gr}(k,n)$
Grassmannian integral at $k{=}1$.
Finally, we present the corresponding
single-minus superamplitude
in $\cN{=}8$ supergravity.

}
\maketitle

\flushbottom
\tableofcontents

\newpage

\section{Introduction}
\label{sec:intro}

Tree-level single-minus gluon amplitudes are usually considered to be trivial, since in Lorentzian signature they vanish identically. However, it was recently observed in \cite{Guevara:2026qzd} that in
$(2,2)$ signature, non-vanishing colour-ordered single-minus amplitudes
$A_n(1^-,\text{rest}^+)$ exist for all multiplicities.
These amplitudes are supported on the half-collinear locus
$\lambda_i{\propto}\lambda_j$ for all $i,j$  and are proportional to
products of delta functions enforcing the collinear condition.

A peculiar feature of these amplitudes is the appearance of a 
``stripped amplitude'' $A_{1\ldots n}$ which is  piecewise constant and built from  
products of sign functions of kinematic invariants.
In twistor space,
tree-level N${}^{q-2}$MHV amplitudes
localise on holomorphic curves of degree $d=q-1$,
where $q$ is the number of negative-helicity gluons~\cite{Witten:2003nn}.
For single-minus amplitudes $d=0$ and 
the curve collapses to a point,
corresponding precisely to the condition
 $\vev{ij}=0$  
that defines the half-collinear
locus~\cite{Witten:2003nn,Roiban:2004yf}.
The first member of this family occurs at three points. Indeed, although three-point amplitudes vanish for real Lorentzian massless kinematics, non-vanishing amplitudes exist in complex and split-signature kinematics,%
\footnote{The need for complex momenta in this context was noted long ago in~\cite{Goroff:1985th}.}
and their explicit form is given by the familiar Parke--Taylor expressions~\cite{Parke:1986gb,Mangano:1987xk}.
These three-point amplitudes were subsequently
used as the fundamental building blocks in the
BCFW on-shell recursion
relations~\cite{Britto:2004ap,Britto:2005fq}.

Before turning to our results, we wish to comment on the distributional nature of these  amplitudes and stress that such contributions are not optional but rather a generic feature of tree amplitudes. They follow directly from the Feynman rules of the theory with the standard $i\varepsilon$ prescription~\cite{Guevara:2026qzd}, and in  half-collinear kinematics, {\it all} split-signature tree amplitudes, including MHV ones, acquire terms of this kind. Usually, tree amplitudes are evaluated at generic kinematics  with the $i\varepsilon$ prescription suppressed, and the results are presented as rational functions.  We will refer to this as the rational  part of the amplitude, while any distributional contributions are omitted. 
  Single-minus amplitudes are distinguished not by being distributional, but by the fact that the rational part of the amplitude, which is supported at generic kinematics, vanishes identically: the distributional contribution is therefore all that remains. Any statement about the symmetries of the $S$-matrix is consequently incomplete unless it also includes this sector.%
\footnote{We return to this point in Section~\ref{sec:analiticity}.}

With these remarks in mind, we turn to 
the central question of this   paper, that  is whether dual superconformal
symmetry -- the hidden symmetry of planar $\cN{=}4$ super Yang--Mills which
underlies its Yangian structure~\cite{Drummond:2009fd} and its conjectured
integrability -- extends to the distributional sector.  This is a
question  of great interest independently of any application
to cross sections: $\cN{=}4$ super Yang--Mills is a theoretical laboratory, and
its hidden symmetries are among the things one most wants to understand about
it.  Our main result is that this defining symmetry does survive in the distributional sector. Note that this sector lies outside the reach of the supersymmetric BCFW recursion originally used to prove dual superconformal symmetry of tree amplitudes~\cite{Brandhuber:2008pf}.  Indeed, BCFW ignores the $i\varepsilon$ prescription and reconstructs only the rational part of the amplitude, whereas the amplitudes studied here are entirely   distributional.

It is also worth recalling in this context  that amplitudes without support on generic real Lorentzian kinematics have been immensely important in the past.  The three-point amplitudes mentioned above exist only in complex or
split-signature kinematics, and are nonetheless the seeds of the BCFW recursion,
and hence of the entire  $S$-matrix.

Having clarified the role of the distributional sector, we now turn to the main subject of this paper: the construction of the $\cN=4$ supersymmetric completion of the single-minus gluon amplitudes for arbitrary multiplicity. Since this sector lies outside the reach of the supersymmetric BCFW recursion, our construction proceeds instead by requiring consistency with the known component amplitudes and the structure of the $\cN{=}4$ supermultiplet. Tree-level $\cN{=}4$ superamplitudes are known to be dual superconformal covariant~\cite{Drummond:2008vq,Brandhuber:2008pf}. However, the standard proof relies on supersymmetric BCFW recursion and therefore does not extend to the amplitudes considered here. Hence, establishing their dual superconformal properties   requires a separate analysis, which we carry out in this paper.
 To  achieve this goal, we will cast  the $n$-point single-minus superamplitude as a product of three building blocks with distinct roles:  
a collinear measure $\Delta^{(n-1)}$, 
which has uniform little-group weight at every leg and absorbs the $\mathbb{Z}_2$ weight of $A_{1\ldots n}$; a stripped amplitude $\tilde{A}_{1\ldots n}$ that is strictly helicity blind and dual conformal invariant; and supersymmetrised momentum conservation delta functions.
The measure $\Delta^{(n-1)}$ is not only cyclic but fully permutation invariant, a property that will be relevant in the extension to $\cN=8$ supergravity.

The single-minus superamplitude we construct 
has  Grassmann degree~four, and 
like its gluonic counterpart in \cite{Guevara:2026qzd} it  is written in terms of an auxiliary reference spinor $|r\rangle$ though it is  independent of its particular choice.
We verify $q$ and  $\bq$ supersymmetry invariance 
at all~$n$, and show that the superamplitude reduces to
the known anti-MHV formula of~\cite{Brandhuber:2008pf,Arkani-Hamed:2008owk} for $n=3$.

After re-expressing $A_{1\ldots n}$ derived in \cite{Guevara:2026qzd} in a  covariant form,  i.e.~without specifying a particular frame, we find that 
$\tilde{A}_{1\ldots n}$  is  invariant
under dual conformal inversion at all~$n$.
The key step is a covariant rewriting
of all sign function arguments as
spinor chains built from differences of  consecutive dual momenta \cite{tHooft:1973alw}
  $P_L{=}x_A-x_B$, $P_R{=}x_B-x_C$, an example  being
$\langle r|\,P_L\,P_R\,|r\rangle$, 
  which transform covariantly
under inversion.
The full superamplitude then 
transforms with weight $\prod_{k=1}^n x_k^2$
under finite dual conformal inversion.
The proof introduces an inverted
reference spinor $|r'\ran \coloneq [r|\,x_1$,
under which each factor of the
superamplitude acquires a definite weight;
the $|r\ran$-independence of the amplitude
then closes the argument.
Combined with supersymmetry and superconformal invariance, this implies invariance under the full  dual 
superconformal group.

We then discuss the 
$\mathrm{Gr}(k,n)$ Grassmannian
integral~\cite{Arkani-Hamed:2009ljj,Arkani-Hamed:2012zlh} at $k=1$, 
performing the explicit localisation. 
Factorising the superamplitude  as
$\cA_n^{(-)} = i^{2-n}\,\tilde{A}_{1\ldots n}\,  F_n$,
where $F_n$ contains no sign functions,   the Grassmannian integral reproduces $F_n$. In other words, the Grassmannian captures only the rational part of the superamplitude, but not $\tilde{A}_{1\ldots n}$. 
Finally we write the corresponding
single-minus superamplitude
in $\cN=8$ supergravity, where the same building blocks appear: $\Delta^{(n-1)}$ and the supersymmetrised delta functions are common to both theories, while the stripped amplitude $\tilde{M}_{1\ldots n}$ is built from absolute values of the same helicity-blind brackets that enter $\tilde{A}_{1\ldots n}$ through sign functions.

The rest of the paper is organised as follows. In Section~\ref{sec:background} we review split-signature kinematics, superspace and dual superspace, and discuss the analyticity properties of the single-minus amplitudes. In particular, we explain why these amplitudes lie outside the reach  of the standard BCFW recursion relations. Section~\ref{sec:superamp} constructs the $\cN{=}4$ superamplitude, verifies its supersymmetry invariance, and matches it to the known anti-MHV superamplitude at $n{=}3$. In Section~\ref{sec:dsc} we prove the dual superconformal covariance of the $n$-point superamplitude. Section~\ref{sec:grassmannian} studies the $k{=}1$ Grassmannian localisation and its relation to the single-minus superamplitude. Section~\ref{sec:sugra} presents the corresponding construction in $\cN{=}8$ supergravity. Finally, in Section~\ref{sec:discussion} we discuss the implications of our results and highlight a number of open questions.

Note: Most calculations in this work were performed with the help of   
{\it Claude Opus 4.6 Extended (Anthropic)} working under supervision of physicists, who confirmed all calculations.

\section{Background}
\label{sec:background}

\subsection{Spinor-helicity variables and
the half-collinear regime}
\label{sec:kinematics}

We use spinor-helicity variables for massless momenta
in split $(2,2)$ signature,%
\footnote{Our conventions are those of \cite{Brandhuber:2022qbk}.}
\be\label{eq:momenta}
p_{i\alpha\dot\alpha}
= \lambda_{i\alpha}\,\lt_{i\dot\alpha}\,,
\ee
where $\lambda_i$ and $\lt_i$ are real and independent, and 
the Lorentz-invariant contractions are
\be\label{eq:brackets}
\vev{ij} = \lambda_i^\alpha\lambda_{j\alpha}\,,
\qquad
\bev{ij} = \lt_{i\dot\alpha}\lt_j^{\dot\alpha}\,, 
\ee
with $2(p_i\Cdot p_j) = \vev{ij} [ji]$.
The {half-collinear regime}~\cite{Guevara:2026qzd}
is defined by
\be\label{eq:halfcoll}
\vev{ij} = 0 \qquad \forall\;i,j\,,
\ee
so that 
\begin{equation}
\label{eq:ci}
    \lambda_i = c_i\lambda\, , 
    \end{equation}
for some common $\lambda$.
Because of the half-collinear kinematics, momentum conservation
$\sum_i p_i = 0$
can be expressed as 
\begin{align}
\label{eq:mom-cons}
  \sum_i \vev{ri}\lt_i=0  
\, , 
\end{align}
where $|r\ran$ is a constant reference spinor
not proportional
to $\lambda_i$,  $\vev{ri}\neq 0$.

\medskip
\noindent
{\it Special frame.}
Following~\cite{Guevara:2026qzd}, it is also useful to consider a
special frame where 
\be\label{eq:frame}
\lambda_i = (1,z_i)\,,\qquad
\lt_i = \omega_i(1,\tilde z_i)\,,\qquad
|r\ran = (0,1)\,.
\ee
In this frame
\be\label{eq:frame_brackets}
\vev{ij} = z_{ji}\,,\qquad
\bev{ij} = \omega_i\omega_j\tilde z_{ij}\,,\qquad
\vev{ri} = -1\,, 
\ee
with $z_{ab} =z_a-z_b$ and $\tilde{z}_{ab} =\tilde{z}_a-\tilde{z}_b$.
In the half-collinear kinematics all the $z_i$ are equal and 
\be\label{eq:ci_unity}
c_i = \frac{\vev{ri}}{\vev{r\lambda}}
= 1 \quad\text{for all }i\,.
\ee
Momentum conservation
$\sum_i p_i = \lambda\sum_i\lt_i = 0$
reduces to $\sum_i\lt_i = 0$.

\medskip
\noindent
{\it The single-minus gluonic amplitude.}
Next we review the $n$-point single-minus gluon amplitude
from~\cite{Guevara:2026qzd}, with particle~$1$ as the
negative-helicity gluon.
Suppressing from now on a factor of $(2\pi)^4$,
this reads~\cite{Guevara:2026qzd}
\be\label{eq:Guevara:2026qzdk}
A_n(1^-,\text{rest}^+) = i^{2-n}
\frac{\vev{r1}^{n+1}}{\prod_{a\neq 1}\vev{ra}}\;
A_{1\ldots n}\;
\prod_{a\neq 1}\delta(\vev{1a})\;
\delta^{(2)}\!\pa{\sum_i\vev{ri}\lt_{i\dot{\alpha}}}\,,
\ee
where the  stripped amplitude
$A_{1\ldots n}$ carries no helicity weight up to possible sign functions. Examples of such functions at $n=3$ and $n=4$ are%
\be\label{eq:A34}
\begin{split}
A_{123} = \sg_{12}  \,,
\qquad
A_{1234} = \tfrac{1}{2}(\sg_{12}\sg_{34}+\sg_{23}\sg_{41})
\,, 
\end{split}
\ee
where
\be
\sg_{ij} \coloneq \sg(\bev{ij})\, . 
\ee
Each sign factor takes values
$+1$, $-1$, or $0$,
so the stripped amplitude is a
piecewise-constant integer.

In the special  frame \eqref{eq:frame}, the  $n$-gluon amplitude
collapses to
\be\label{eq:gluon_frame}
A_n(1^-,\text{rest}^+) = i^{2-n}\,A_{1\ldots n}\,
\prod_{a\neq 1}\delta(z_{1a})\;
\delta^{(2)}\!\pa{\sum_i\lt_i}\, . 
\ee
Although particle~$1$ is singled out in the product
of $\delta (z_{1a})$, the support $z_{1a}=0$
for all~$a$ implies $z_{ij}=0$ for all pairs,
so any particle could equivalently be chosen.
 For $n\geq 5$,  composite sign functions
such as $\sg_{1,23} = \sg(\bev{12}+\bev{13})$
also appear \cite{Guevara:2026qzd} for expressions of $A_{1\ldots n}$ derived in the special frame. 
It is clear that objects like $\bev{12}+\bev{13}$ do not  transform appropriately under the little group; 
in Section~\ref{sec:sign_functions}
we express  all sign-function arguments
in a covariant form%
\footnote{This possibility was also mentioned in \cite{Guevara:2026qzd} and turns out to be important for the proof of dual conformal invariance presented in Section~\ref{sec:sign_functions}.}
as spinor chains, e.g.~$\langle r|\,P_L\,P_R\,|r\rangle$, 
built from differences of consecutive dual momenta. Such spinor chains are helicity blind and, importantly, have transparent  dual
conformal transformation properties.

\subsection{Comments on analyticity properties and the BCFW recursion} 
\label{sec:analiticity}

The piecewise-constant dependence of $A_{1\ldots n}$ on the kinematics is not in conflict with analyticity. In complex Minkowski space when considering tree amplitudes the $i\varepsilon$-prescription is usually ignored
and they are treated as ordinary meromorphic functions with simple poles such as $1/s$ and no distributional contributions.

The familiar Plemelj formula, 
\begin{align} \frac{1}{s+i\varepsilon} = \mathrm{PV}\,\frac{1}{s} -i\pi\,\delta(s), 
\end{align} 
arises only when one restricts to a real slice and interprets the propagator as a distribution.  Equivalently, the distinction between principal-value and $\delta$-function terms appears only after taking an appropriate real boundary value. This is precisely why the familiar tree-level statement $A_n(1^-,2^+,\ldots,n^+)=0$ holds literally in complex kinematics: the amplitude vanishes as a meromorphic function. The non-vanishing single-minus amplitudes encountered in split signature are invisible from this perspective: they arise only after restricting to a real slice, where meromorphic singularities such as $1/\langle ij\rangle$ acquire distributional support. 

Distributional terms of this sort are a generic feature of amplitudes, but they are normally obscured by the meromorphic part, which has support at generic kinematics. Single-minus amplitudes are special because their meromorphic part vanishes identically at tree level and the amplitude is entirely captured by the distributional contribution. The same phenomenon occurs at half-collinear kinematics in split signature for other amplitudes,  including MHV amplitudes. In those cases the distributional contribution coexists with a non-vanishing meromorphic term and is therefore less conspicuous (and usually ignored).

This observation has an important consequence for on-shell recursion relations. 
The standard BCFW recursion treats the amplitude as a meromorphic function and discards the $i\varepsilon$ prescription, so it is sensitive only to the rational part of the amplitude.  Single-minus amplitudes, by contrast, have vanishing rational part and are entirely distributional.
 The contour argument requires the shifted propagators to vanish at isolated values of the deformation parameter, so that the deformed amplitude has simple poles on which the recursion localises. On the half-collinear support this fails outright: since all angle brackets vanish there, the shifted propagators vanish identically in the deformation parameter rather than at isolated points, so there is no pole to localise on and the derivation does not get off the ground. 
 Consequently, the standard BCFW construction fails to produce any distributional terms  of amplitudes.
 Instead, retaining the $i\varepsilon$ prescription, the propagators produce precisely the delta functions and sign functions that appear in the answer.

\subsection{Superspace and dual superspace}
\label{sec:superspace}

In $\cN=4$ supersymmetric Yang-Mills, amplitudes with a fixed number
of external particles and a given total helicity
are efficiently packaged into
superamplitudes~\cite{Nair:1988bq}.
The on-shell superspace associates to each
particle~$i$ a momentum
$p_i = \lambda_i\lt_i$
and Grassmann variables $\eta_i^A$, $A=1,\ldots,4$.
Expanding the superamplitude in powers
of the $\eta_i^A$,
a monomial carrying $m_i$ powers of~$\eta_i$
describes particle~$i$ with helicity
$h_i = 1 - m_i/2$.
The supercharges are
\be\label{eq:supercharges}
q^{A\alpha } = \sum_i\lambda_{i}^{\alpha}\,\eta_i^A\,,
\qquad
\bq_{A\dot\alpha} = \sum_i\lt_{i\dot{\alpha}}\,
\frac{\partial}{\partial\eta_i^A}\,.
\ee
The tree-level $n$-point  MHV superamplitude is given by~\cite{Nair:1988bq}
\be\label{eq:Nair}
\cA_{\mathrm{MHV}}(1,\ldots,n)
= i\,
\frac{\delta^{(4)}(p)\;
\delta^{(8)}(q)}
{\vev{12}\vev{23}\cdots\vev{n1}}\,,
\ee
where $p \coloneq \sum_i\lambda_i\lt_i$
and $q\coloneq\sum_i\lambda_i\eta_i$
are the total momentum and supermomentum,
respectively.
Any tree-level superamplitude must be
annihilated by both supersymmetry charges, 
\be\label{eq:susy_ward}
q^{A\alpha}\,\cA_n = 0\,,\qquad
\bq_{A\dot\alpha}\,\cA_n = 0\,.
\ee
In order to study dual superconformal symmetry, we introduce dual superspace
variables and their transformation
properties under dual conformal inversion \cite{Drummond:2008vq}.
In particular
\be\label{eq:dualdef}
(x_i - x_{i+1})^{\dot\alpha\alpha}\coloneq p_i^{\dot\alpha \alpha}
= \lambda_i^\alpha\lt_i^{\dot\alpha}\,,
\qquad
(\theta_i - \theta_{i+1})^{A\alpha}\coloneq q^{A\alpha}_i 
= \lambda_i^\alpha\eta_i^A\,,
\ee
with $x_{n+1}=x_1$ and  $\theta_{n+1}=\theta_1$.
Dual conformal inversions act as~\cite{Drummond:2008vq}
\be\label{eq:inversions}
\begin{split}
I[x_{i}] &= x_i^{-1}\,,\qquad \quad \qquad \quad I[\theta_i^{A\alpha}]
= (x_i^{-1})^{\dot\alpha\beta}\theta_{i,\beta}^A\,, \\
I[\lambda_i^\alpha]
&= (x_i^{-1})^{\dot\alpha\beta}\lambda_{i \beta}\,,\qquad \quad 
\ I[\lt_i^{\dot{\alpha}}]= -\lt_{i \dot{\beta}} (x_{i+1}^{-1})^{\dot{\beta} \alpha}\, , 
\end{split}
\ee
and spinor brackets transform as
\be\label{eq:bracket_inv}
I[\vev{i\;i{+}1}] = \frac{\vev{i\;i{+}1}}{x_i^2}\,,
\qquad
I[\bev{i\;i{+}1}] = \frac{\bev{i\;i{+}1}}{x_{i+2}^2}\,.
\ee
Note that because of the half-collinear kinematics we can equivalently  write
\begin{align}
    I[\lambda_i^\alpha]
= \frac{x_1^{\dot\alpha\beta}}{x_i^2}\lambda_{i\beta}\, , 
\end{align}
since $(x_1-x_i)^{\dot\alpha\alpha}\lambda_{i\alpha}=0$. 

Dual superconformal covariance is  the statement that any tree-level superamplitude satisfies~\cite{Drummond:2008vq,Brandhuber:2008pf}
\be\label{eq:expected_weight}
I[\cA_n] = \prod_{k=1}^n x_k^2\;\cA_n\,.
\ee
This was conjectured in \cite{Drummond:2008vq} and shown to hold in several examples, and later proved in \cite{Brandhuber:2008pf} using a supersymmetric extension of the BCFW recursion relation. 
Because of the half-collinear kinematics the infinite sequence of single-minus superamplitudes cannot be derived from this recursion relation, and we will provide  an alternative proof of dual superconformal invariance in Section~\ref{sec:dsc}.

The simplest member of the infinite sequence of single-minus superamplitudes  is the three-point anti-MHV superamplitude \cite{Brandhuber:2008pf,Arkani-Hamed:2008owk},
whose dual superconformal covariance
was established in~\cite{Brandhuber:2008pf}. Explicitly, it is given by 
\be\label{eq:Brandhuber:2008pf}
\cA_{\overline{\mathrm{MHV}}}(1,2,3)
= -i\,
\frac{\delta^{(4)}(P)\;\delta^{(4)}
(\eta_1\bev{23}+\eta_2\bev{31}+\eta_3\bev{12})}
{\bev{12}\bev{23}\bev{31}}\,.
\ee
This expression can also be obtained from the
Nair MHV superamplitude~\cite{Nair:1988bq}
by performing a Grassmann Fourier transform combined with
the exchange
$\lambda\leftrightarrow\lt$~\cite{Arkani-Hamed:2008owk}.
The single-minus superamplitude constructed
in the next section reduces
to~\eqref{eq:Brandhuber:2008pf} at $n=3$;
this is the only multiplicity at which
the result can be recast in a form
that is manifestly independent
of the reference spinor~$|r\rangle$.

\section{The \texorpdfstring{$\mathcal N=4$}{N=4} superamplitude}
\label{sec:superamp}

\subsection{Construction of the superamplitude}
\label{sec:definition}

We seek the $\cN=4$ superamplitude whose
$(\eta_k)^4$ component reproduces the gluon
amplitude~\eqref{eq:gluon_frame} for each~$k$.
On the collinear support $\lambda_i = c_i\lambda$,
the supermomentum has rank~1:
\be\label{eq:rank1}
q^A_\alpha = \sum_i\lambda_{i\alpha}\,\eta_i^A
= \lambda_\alpha\sum_i c_i\, \eta_i^A\,,
\ee
so only four of the eight components of $q^A_\alpha$
are independent.
Supermomentum conservation therefore
imposes only four Grassmann conditions  which, paralleling \eqref{eq:mom-cons}, can be expressed as 
\begin{equation}
    \sum_i \vev{ri}\eta^A_i=0\, . 
\end{equation}
The 
 natural fermionic delta function is then $\delta^{(4)}(\sum_i\vev{ri}\eta_i) =\delta^{(4)}(\sum c_i\eta_i)$.
In the special frame where $c_i=1$ it becomes   
$\delta^{(4)}(\sum\eta_i)$, which 
gives coefficient~one for $(\eta_k)^4$ for every~$k$,
 as required.
Every other Grassmann component also has
unit coefficient,
so all component amplitudes (gluons, gluinos, scalars)
are equal on the collinear support in that frame.

These observations determine the structure of  the superamplitude
uniquely.
For any choice of the reference spinor $|r\ran$
not proportional to $\lambda$, we claim that the superamplitude 
takes
the form
\be\label{eq:superamp}
\cA_n^{(-)} = \;i^{2-n}\;
\tilde{A}_{1\cdots n}(\tilde\Lambda) \;\Delta^{(n-1)} (\{\lambda_i\})\;
\delta^{(2)}\!\pa{\sum_i\vev{ri}\lt_{i\dot{\alpha}}}\;
\delta^{(4)}\!\pa{\sum_i\vev{ri}\eta_i^A}\,,
\ee
where we now describe all the building blocks. 

{\bf 1.} 
The first building block  $\delta^{(2)}\!\pa{\sum_i\vev{ri}\lt_{i\dot{\alpha}}}\;
\delta^{(4)}\!\pa{\sum_i\vev{ri}\eta_i^A}$ is the natural supersymmetrisation of the bosonic delta function $\delta^{(2)}\!\pa{\sum_i\vev{ri}\lt_{i\dot{\alpha}}}$.

{\bf 2.} 
The second  is a product of $(n{-}1)$ delta functions imposing the half-collinear kinematics,  
\be \label{eq:Delta}
\Delta^{(n-1)} (\{\lambda_i\})\coloneq 
\frac{\prod_{a=1}^{n-1}\delta(\vev{a\;a{+}1})}
{|\vev{rn}\vev{r1}|}\, ,
\ee
which has uniform helicity weight (including signs) for every particle. Note that  the delta functions appearing in \eqref{eq:Delta} are made of brackets  containing only adjacent spinors (in colour space), which will prove particularly useful for studying dual conformal properties. Furthermore, it is not only cyclic invariant, but also invariant under permutations, as we will now show. This will be relevant in the construction of the $\cN=8$ supergravity amplitude in Section~\ref{sec:sugra}.

In order to check permutation invariance of $\Delta^{(n-1)}$, it suffices to check invariance under the swap of particles $i$ and $i{+}1$
since adjacent transpositions $(i,i{+}1)$ for $i=1,\ldots,n{-}1$ generate $S_n$.   On the support of $\delta(\vev{i\;i{+}1})$ we have $\vev{i\;i{+}1}=0$, so by Schouten%
\footnote{The change of variables in \eqref{eq:schouten35} is understood in the distributional sense, on the support of the spectator factor $\delta(\vev{i\,i{+}1})$. Since $\vev{r\,i}\neq0$ by construction, the corresponding Jacobian is non-singular.}

 \be\label{eq:schouten35}
\vev{i{-}1\;i{+}1} = \frac{\vev{r\;i{+}1}}{\vev{r\,i}}\vev{i{-}1\;i}\,, \qquad
\vev{i\;i{+}2} = \frac{\vev{r\,i}}{\vev{r\;i{+}1}}\vev{i{+}1\;i{+}2}\,.
\ee
For an interior swap ($2{\leq} i{\leq} n{-}2$), the denominator $|\vev{r1}\vev{rn}|$ is unchanged. The only affected delta functions are $\delta(\vev{i{-}1\;i}){\to}\delta(\vev{i{-}1\;i{+}1})$ and $\delta(\vev{i{+}1\;i{+}2}){\to}\delta(\vev{i\;i{+}2})$, 
picking up Jacobians $|\vev{ri}/\vev{r\;i{+}1}|$ and $|\vev{r\;i{+}1}/\vev{ri}|$ respectively, which cancel.
For the boundary swap $(1,2)$, only $\delta(\vev{23})\to\delta(\vev{13})$ is affected, with Jacobian $|\vev{r2}/\vev{r1}|$. The denominator changes from $|\vev{r1}\vev{rn}|$ to $|\vev{r2}\vev{rn}|$, contributing a factor $|\vev{r1}/\vev{r2}|$, which cancels the Jacobian. The swap $(n{-}1,n)$ is analogous.

{\bf 3.} The function $\tilde{A}_{1\ldots n}$ is a cyclic invariant, 
helicity-blind function (including signs)
which can be obtained from the expression $A_{1\ldots n}$ introduced in \cite{Guevara:2026qzd} with an appropriate covariantisation procedure:  
\be
\label{eq:rep1}
\sg_{L,R} \coloneq \sg [\tilde\lambda_L \tilde\lambda_R] \to \sg [\tilde\Lambda_L \tilde\Lambda_R]\, , 
\ee
where
\begin{align}
    \tilde\lambda_A &\coloneq \sum_{i\in A}\tilde\lambda_i \,  , 
    \\ 
    \label{eq:tLam}
    \tLam_A &\coloneq  \sum_{i\in A}\vev{ri}\lt_i\, , 
\end{align}
and where the sum involves consecutive labels. For instance, using the results   of \cite{Guevara:2026qzd}
\be
\begin{split}
A_{123} &= \sg_{12}\, , \\
A_{1234} &= \frac{1}{2}\big( \sg_{12}\sg_{34} + \sg_{23}\sg_{41}\big)\, , 
\end{split}
\ee
we have 
\begin{align}
\label{eq:3ptAtfinal}
\tilde{A}_{123} &= \sg (\vev{r1}[12]\vev{r2}) = - \sg \langle r| 12 |r\rangle\, ,
\end{align}
and
\begin{align}
\label{eq:4ptAtfinal}
\begin{split}
\tilde{A}_{1234}  & = \frac{1}{2}\Big[ \sg (\vev{r1} \vev{r2}[12] \vev{r3} \vev{r4}[34] ) +
\sg (\vev{r2} \vev{r3}[23] \vev{r4} \vev{r1}[41] )\Big] \\ 
& = 
\frac{1}{2}\Big[ \sg (\langle r| 12|r \rangle  \langle r| 34|r \rangle) + 
\sg( \langle r| 23|r \rangle  \langle r| 41|r  \rangle)
\Big]
\, . 
\end{split}
\end{align}
Note that  $\tilde{A}_{123}$ and $\tilde{A}_{1234}$ are both cyclic invariant. Furthermore,  in the frame \eqref{eq:frame} we have $\vev{ri} = -1$ and hence $\tLam_A = -\tilde{\lambda}_A$.
The blocks $L$ and  $R$ are also consecutive in 
colour ordering.
The fact that these blocks are made of consecutive momenta has important consequences from the point of view of dual conformal symmetry,  as we will show in Section~\ref{sec:sign_functions}. Further properties of the superamplitude will be discussed in Section \ref{sec:furtherproperties}.

\medskip
\noindent
{\it Supersymmetry invariance.}
It is immediate to see that the superamplitude satisfies
the supersymmetry Ward
identities~\eqref{eq:susy_ward}. Specifically, 
$q$-supersymmetry is manifest, 
while  for $\bq$,  acting with $\partial/ \partial \eta$ on
$\delta^{(4)}(\sum_i \vev{ri}\eta_i)$
gives  a 
result  proportional to
$\sum_i\vev{ri}\lt_{i\dot{\alpha}}$,
which vanishes on the support of
the other delta function $\delta^{(2)}(\sum_i \vev{ri}\lt_i)$.
The stripped amplitude $\tilde{A}_{1\ldots n}$ is transparent
to both generators, having no $\eta$-dependence.
Furthermore, 
$\delta^{(4)}(\sum_i\vev{ri}\eta_i^A)$
is the unique $SU(4)$-invariant Grassmann polynomial of degree~four
that  enforces supermomentum conservation
on the half-collinear support.

Summarising, \eqref{eq:superamp} is the unique
$\cN{=}4$ supersymmetric completion of the
single-minus gluon amplitude~\eqref{eq:Guevara:2026qzdk}.
The explicit match with the anti-MHV formula
of~\cite{Brandhuber:2008pf,Arkani-Hamed:2008owk} at $n{=}3$
is verified in Section~\ref{sec:3pt}, and 
in Section~\ref{sec:dsc} we further show that
the superamplitude is dual superconformal covariant
for arbitrary $n$.
In the next section we show that the gluonic component of our superamplitude agrees with  the expression of \cite{Guevara:2026qzd}. 
A similar supersymmetrisation will hold in $\cN=8$ supergravity, 
 \begin{align}
\delta^{(2)}\!\pa{\sum_i\vev{ri}\lt_{i\dot{\alpha}}}\;
\to \delta^{(2)}\!\pa{\sum_i\vev{ri}\lt_{i\dot{\alpha}}}\;
\delta^{(8)}\!\pa{\sum_i \vev{ri}\eta_i^A}\, , 
\end{align}
where now $A{=}1, \ldots , 8$.  This  will be employed in Section~\ref{sec:sugra} for constructing the 
$\cN{=}8$ supersymmetric completion of the single-minus graviton amplitudes of \cite{Guevara:2026qwa}.

\subsection{Comparison to single-minus gluon amplitude}

The supersymmetric amplitude \eqref{eq:superamp} contains the single-minus amplitude discussed in \cite{Guevara:2026qzd} in its components, and we will show this explicitly below.

To extract the component amplitude where all external particles are positive-helicity gluons and particle $1$ is a negative-helicity gluon, we need to pick the term proportional to $(\eta_1)^4$. This gives
\be\label{eq:singleminusadj}
A_n(1^-,\text{rest}^+) = i^{2-n}\,
\frac{\vev{r1}^4}{|\vev{r1}\vev{rn}|}\;
\tilde{A}_{1\cdots n}\;
\prod_{a=1}^{n-1}\delta(\vev{a\;a{+}1})\;
\delta^{(2)}\!\pa{\sum_i\vev{ri}\lt_{i\dot{\alpha}}}\,.
\ee
Then we need to change basis for the spinor delta functions, from $\prod_{a=1}^{n-1}\delta(\vev{a \, a+1})$ to $\prod_{a=2}^{n}\delta(\vev{1 \, a})$. The Schouten identity gives,
for each $a=1,\ldots,n{-}1$,
\be\label{eq:schouten_delta}
\vev{r1}\,\vev{a\;a{+}1}
= \vev{ra}\,\vev{1,a{+}1}
- \vev{r,a{+}1}\,\vev{1a}\,,
\ee
which defines an $(n{-}1){\times}(n{-}1)$ matrix~$M$
via $\vev{r1}\,\vev{a\;a{+}1}
= \sum_{b\neq 1}M_{ab}\,\vev{1b}$.
The matrix $M$ is an $(n{-}1)\times(n{-}1)$
lower bidiagonal matrix
with diagonal entries $\vev{ra}$
for $a=1,\ldots,n{-}1$.
The determinant is the product
of diagonal entries,
$\det M
= \vev{r1}
\prod_{a=2}^{n-1}\vev{ra}$.
Since the left-hand side
of~\eqref{eq:schouten_delta}
carries a factor $\vev{r1}$,
the Jacobian of the change of variables
$\{\vev{1a}\}_{a\neq 1}
\to\{\vev{a\;a{+}1}\}_{a=1}^{n-1}$
is $\det M/\vev{r1}^{n-1}
= \prod_{a=2}^{n-1}\vev{ra}
\,/\,\vev{r1}^{n-2}$,
giving
\be\label{eq:delta_identity}
\prod_{a=1}^{n-1}\delta(\vev{a\;a{+}1})
=
\frac{|\vev{r1}|^{n-2}}{\prod_{a=2}^{n-1}|\vev{ra}|}\; \prod_{a=2 }^n\delta(\vev{1a})
\,.
\ee
Combining with the prefactor
in~\eqref{eq:singleminusadj}, we get
\be\label{eq:kindep_proof}
\frac{\vev{r1}^4}{|\vev{r1}\,\vev{rn}|} \frac{|\vev{r1}|^{n-2}}{\prod_{a=2}^{n-1}|\vev{ra}|} 
= \frac{\vev{r1}^{n+1}}{\prod_{a\neq 1}\vev{ra}} \sg \Big(\vev{r1}^{n+1} \prod_{a=2}^n\vev{ra} \Big)
\,.
\ee
Therefore the  superamplitude   takes
the form
\be\label{eq:superamp-old-old}
\cA_n^{(-)} = i^{2-n}
\frac{\vev{r1}^{n-3}}{\prod_{a\neq 1}\vev{ra}}\;
A_{1\cdots n}\;
\prod_{a=2}^{n}\delta(\vev{1a})\;
\delta^{(2)}\!\pa{\sum_i\vev{ri}\lt_{i\dot{\alpha}}}\;
\delta^{(4)}\!\pa{\sum_i\vev{ri}\eta_i^A}\,,
\ee
where
\be
\label{eq:tildeA}
{A}_{1\ldots n}
= \tilde{A}_{1\ldots n}\,
\sg \Big(\vev{r1}^{n} \prod_{a=1}^n \vev{ra} \Big)\, .
\ee
Note that on the collinear support
this sign is $\sg(c_1^{n}\prod_{a=1}^n c_a)$,
which is independent of the reference
spinor~$|r\ran$: it is the residual $\mathbb{Z}_2$ helicity weight carried by $A_{1\ldots n}$ in a general frame, which the collinear measure $\Delta^{(n-1)}$ absorbs when passing to the helicity-blind $\tilde{A}_{1\ldots n}$. This weight is invisible in the special frame of~\cite{Guevara:2026qzd}, where $c_i=1$ (equivalently $\vev{ri}=-1$) for all~$i$ and the sign becomes trivial: there, $\tilde{A}_{1\ldots n}=A_{1\ldots n}$ and $\sg[\tilde\Lambda_A\tilde\Lambda_B]=\sg[\tilde\lambda_A\tilde\lambda_B]$, confirming that the result~\eqref{eq:superamp} agrees with the single-minus amplitude of~\cite{Guevara:2026qzd}, presented in that frame~\eqref{eq:frame}.

\subsection[Agreement with the three-point anti-MHV amplitude]{Agreement with the three-point $\overline{\mathrm{MHV}}$ amplitude}
\label{sec:3pt}

For $n{=}3$, the superamplitude~\eqref{eq:superamp}
must reduce to the well-known anti-MHV
superamplitude~\eqref{eq:Brandhuber:2008pf} of~\cite{Brandhuber:2008pf,Arkani-Hamed:2008owk}.
At three points, we can  write%
\footnote{It is well known that for three massless particles the half-collinear condition follows from momentum conservation. In this sense, the single-minus amplitude can be viewed as a natural higher-point generalisation of the much better-known three-point amplitude.}
\begin{align}\label{eq:d4_adj_cov}
\delta^{(4)} (\sum_{i=1}^3 \lambda_i\lt_i) = |J| \, \delta (\vev{12})\delta (\vev{23})  \delta^{(2)} (\sum_{i=1}^3 \vev{ri}\lt_i) \, , 
\end{align}
with 
\begin{align}\label{eq:J_cov}
   |J| = \frac{|\vev{r1}|\,|\vev{r2}|}{|\bev{23}|}\, . 
\end{align}
The  Jacobian in~\eqref{eq:J_cov} can be derived 
as follows. 
Writing $\lambda_i = c_i\lambda + \epsilon_i\mu$
with $\vev{\lambda\mu}=1$,
the change of variables from
$(P^{1\dot 1}, P^{1\dot 2},
P^{2\dot 1}, P^{2\dot 2})$
to
$(\vev{12},\;\vev{23},\;
f^{\dot 1},\;f^{\dot 2})$,
where
$f^{\dot\alpha}
\coloneq \sum_{i=1}^3\vev{ri}\lt_i^{\dot\alpha}$,
has Jacobian 
\be\label{eq:adj_jacobian}
\frac{\partial(\vev{12},\vev{23},
      f^{\dot 1},f^{\dot 2})}
     {\partial(P^{1\dot 1},P^{1\dot 2},
      P^{2\dot 1},P^{2\dot 2})}
= \frac{c_1\,c_2\,\vev{r\lambda}^2}
       {\bev{23}}\,.
\ee
Since $c_i\vev{r\lambda}=\vev{ri}$, this immediately gives \eqref{eq:J_cov}.

Hence, the three-point superamplitude  can  be recast in the form 
\begin{align}
  &\cA_{3, \overline{\rm MHV}} = -i\, \frac{\delta^{(4)} (\sum_{i=1}^3 \lambda_i\lt_i)\delta^{(4)} ( \eta_1 [23] + \eta_2[31] + \eta_3[12])}{[12][23][31]} \\ & = -i\frac{|\vev{r1}\vev{r2}|}{[12][23]^2[31]}\sg_{23} \delta (\vev{12})\delta (\vev{23})\, \delta^{(2)} (\sum_{i=1}^3 \vev{ri}\lt_i) \, \delta^{(4)} ( \eta_1 [23] + \eta_2[31] + \eta_3[12])\, . \nonumber
\end{align}
The idea now is to rewrite everything in terms of angle brackets. Using the three-point identities  
  \begin{align}
    [31] =\frac{\vev{r2}}{\vev{r1}}[23]\, , \qquad  
  [12] = \frac{\vev{r3}}{\vev{r1}}[23]\, ,
  \end{align}
    we arrive at
  \begin{align}
  \label{eq:3-pt-super-deltas}
\begin{split}
  \cA_{3, \overline{\rm MHV}} =   -i\sg(\vev{r2}\vev{r3})\sg_{23} \frac{\delta (\vev{12})\delta (\vev{23})}{|\vev{r1}\vev{r3}|}\delta^{(2)} (\sum_{i=1}^3 \vev{ri}\lt_i)\, 
  \delta^{(4)} \Big( \sum_{i=1}^3 \vev{ri}\eta_i\Big)
  \, . 
\end{split}
\end{align}
 Comparing with \eqref{eq:superamp} at $n=3$ we see that 
 \begin{align}
\label{eq:Atilde123}     
\tilde{A}_{123}=\sg(\vev{r2}\vev{r3})\sg_{23}\, , 
 \end{align}
 which is equal to the result \eqref{eq:3ptAtfinal} using momentum conservation.

\subsection{Further properties of the superamplitude}\label{sec:furtherproperties}

\medskip
\noindent
{\it Little-group scaling.}
We now verify that~\eqref{eq:superamp}
satisfies the little-group scaling equation~\cite{Witten:2003nn}
\be\label{eq:helicity_op}
h_i\,\cA_n^{(-)}=0
\qquad\text{for all }i\,,
\ee
where
\begin{align}
h_i = \lambda_i\cdot\partial_{\lambda_i}
- \lt_i\cdot\partial_{\lt_i}
- \eta_i\cdot\partial_{\eta_i} + 2\, . 
\end{align}
This is equivalent to requiring that
the little-group transformation 
\be(\lambda_i,\lt_i,\eta_i)
\to(t\lambda_i,t^{-1}\lt_i,t^{-1}\eta_i)\ee
scales the superamplitude by $t^{-2}$.
A  simplification is that
both delta functions in~\eqref{eq:superamp}
are {individually invariant}
under the little group at every leg.
The scaling therefore comes entirely from
the prefactor and collinear deltas.

Under the little-group rescaling
each collinear delta function transforms as
\be\delta(\vev{a\;a{+}1})\to
|t_a\,t_{a+1}|^{-1}\,\delta(\vev{a\;a{+}1})\, . 
\ee
In the product $\prod_{a=1}^{n-1}\delta(\vev{a\;a{+}1})$
each particle~$i$ with $2\leq i\leq n{-}1$
appears in exactly two adjacent delta functions,
$\delta(\vev{i{-}1,\;i})$ and $\delta(\vev{i,\;i{+}1})$,
and therefore contributes $|t_i|^{-2}$.
By contrast, particles~1 and~$n$ sit at the
boundary of the chain, 
contributing $|t_1|^{-1}$
and  $|t_n|^{-1}$.
The denominator $|\vev{rn}\vev{r1}|$
supplies the missing power at each boundary:
$|\vev{r1}|\to|t_1|\,|\vev{r1}|$ and
$|\vev{rn}|\to|t_n|\,|\vev{rn}|$.
Collecting all factors,
\be
\Delta^{(n-1)}
\;\longrightarrow\;
\prod_{i=1}^n t_i^{-2}\;\Delta^{(n-1)}\,.
\ee
Finally we note that the function $\tilde{A}_{1\ldots n}$ as defined in Section~\ref{sec:definition} (see in particular \eqref{eq:rep1} and \eqref{eq:tLam})
carries strictly no little-group weights. 
In conclusion,  every particle carries the same little-group weight, as required.

\medskip\noindent
{\it Independence of $|r\ran$.}
Any change of reference spinor can be written as
$|r\ran\to|r'\ran = a\,|r\ran + b\,|\lambda\ran$,
since $|r\ran$ and $|\lambda\ran$ form a basis.
On the collinear support $\lambda_i = c_i\lambda$,
we have $\vev{\lambda\, i} = 0$ for all~$i$,
so $\vev{r'i} = a\,\vev{ri}$
and the $|\lambda\ran$ component
drops out entirely.
Each factor in~\eqref{eq:superamp}
then transforms as follows.
The function $\tilde{A}_{1\ldots n}$
is $|r\ran$-independent:
from~\eqref{eq:tLam},
$\tLam_A \to a\,\tLam_A$,
and since $\tilde{A}$ is built from
sign functions of brackets
$[\tLam_L,\tLam_R]$
(cf.~\eqref{eq:rep1}),
the common factor $a^2$ drops out
of every $\sg$.
The collinear measure $\Delta^{(n-1)}$
has $|r\ran$-weight~$-2$:
the numerator
$\prod\delta(\vev{a\;a{+}1})$
involves only particle spinors and is
$|r\ran$-independent,
while the denominator
$|\vev{r1}\vev{rn}|
\to a^2\,|\vev{r1}\vev{rn}|$.
The bosonic delta
$\delta^{(2)}(\sum_i\vev{ri}\lt_i)$
contributes weight~$-2$.
Finally, the fermionic delta
$\delta^{(4)}(\sum_i\vev{ri}\eta_i^A)$
is a degree-four Grassmann polynomial
with each factor linear in $\vev{ri}$,
contributing weight~$+4$.
The total weight is zero, 
confirming $|r\ran$-independence.

\medskip
\noindent
{\it Cyclic invariance.}
Cyclic invariance of $\Delta^{(n-1)}$ is a special case of the permutation invariance proved in Section~\ref{sec:definition}. 
The delta functions in~\eqref{eq:superamp} are manifestly permutation invariant and
$\tilde{A}_{1\ldots n}$ is cyclic invariant,
 hence the superamplitude is cyclic invariant.

\medskip
\noindent
{\it Special frame.}
In the frame~\eqref{eq:frame}
($\vev{ri}=-1$, $c_i=1$),
the superamplitude~\eqref{eq:superamp} reduces to
\be\label{eq:superamp-frame}
\cA_n^{(-)} = i^{2-n}\,A_{1\ldots n}\,
\prod_{a=2}^n\delta(z_{1a})\;
\delta^{(2)}\!\pa{\sum_i\lt_i}\;
\delta^{(4)}\!\pa{\sum_i\eta_i}\,.
\ee

\section{Dual conformal analysis}
\label{sec:dsc}

\subsection{Dual conformal inversion}
\label{sec:ferm_inv}

We now establish dual conformal covariance
of the full superamplitude \eqref{eq:superamp}
by finite inversion.
Dual conformal boosts can be decomposed
as $K^\mu = I P^\mu I$, where $I$ is a dual conformal inversion. 
Since the superamplitude is manifestly
invariant under dual translations
$x_i\to x_i + a$, $\theta_i\to\theta_i$,
it suffices to establish covariance
under  inversions.
The key idea is to allow the reference
spinor~$|r\ran$ to transform
under inversion in a specific way discussed  later in~\eqref{eq:rhat1} and \eqref{eq:rhat2};
since the amplitude
is $|r\ran$-independent,
the inverted amplitude with the new
reference spinor $|r'\ran$ equals
the original.

\medskip
\noindent
{\it The superamplitude in dual-space form.}
On the collinear support, the dual superspace
relations~\eqref{eq:dualdef}
give 
\be\label{eq:telescope}
\sum_i\vev{ri}\lt_{i}^{\dot\alpha}
= \lan r|_\alpha(x_1{-}x_{n+1})^{\alpha\dot\alpha}\,,
\qquad
\sum_i\vev{ri}\eta_i^A
= \lan r|_\alpha(\theta_1{-}\theta_{n+1})^{A\alpha}\,.
\ee
Using the cyclic rewriting of the collinear
delta functions, the superamplitude takes
the form
\be\label{eq:cyclic_form}
\cA_n^{(-)} = i^{2-n}\, \tilde{A}_{1\cdots n}\;\frac{\prod_{a=1}^{n-1}\delta(\vev{a\;a{+}1})}{|\vev{rn}\vev{r1}|}\;
\delta^{(2)}\!\pa{\lan r|(x_1{-}x_{n+1})}\;
\delta^{(4)}\!\pa{\lan r|(\theta_1{-}\theta_{n+1})}\,.
\ee

\medskip
\noindent
{\it Inversion of
$\delta^{(2)}(\lan r|(x_1{-}x_{n+1}))$.}
Since inversion maps undotted to dotted spinors, we  define the inverted reference spinor by
\be\label{eq:rhat1}
I(\lan r|) =: [r|\,,
\ee
and also introduce the  auxiliary spinor 
\be\label{eq:rhat2}
\lan r'| \coloneq [r|\,x_1\, .
\ee
The spinor $|r'\ran$ will then play the role of the
reference spinor in the inverted amplitude.
Under inversion
$x_i\to x_i^{-1}$, we then have 
\be
\lan r|(x_1{-}x_{n+1})
\;\longrightarrow\;
[r|(x_1^{-1}{-}x_{n+1}^{-1})
= -[r|\,x_1^{-1}(x_1{-}x_{n+1})x_{n+1}^{-1}\,.
\ee
Using $[r|\,x_1^{-1} = \lan r'|/x_1^2$
and $x_{n+1}^{-1} = x_{n+1}/x_{n+1}^2$,
this becomes
\be\label{eq:I_delta2}
-\frac{1}{x_1^2\,x_{n+1}^2}\,
\lan r'|(x_1{-}x_{n+1})\,x_{n+1}\,.
\ee
The argument of the delta function undergoes a
transformation 
with determinant
\be
\det\!\left[-\frac{x_{n+1}}{x_1^2\,x_{n+1}^2}\right]
= \frac{x_{n+1}^2}{(x_1^2)^2\,(x_{n+1}^2)^2}
= \frac{1}{(x_1^2)^2\,x_{n+1}^2}\,, 
\ee
and hence 
\be
I\!\left[\delta^{(2)}\!\pa{
\lan r|(x_1{-}x_{n+1})}
\right]
= (x_1^2)^2\,x_{n+1}^2\;
\delta^{(2)}\!\pa{
\lan r'|(x_1{-}x_{n+1})}\,.
\ee
On the support of the  momentum-conserving delta functions we can use 
$x_{n+1}=x_1$, hence
\be\label{eq:delta2_weight}
I\!\left[\delta^{(2)}\!\pa{
\lan r|(x_1{-}x_{n+1})}
\right]
= (x_1^2)^3\;
\delta^{(2)}\!\pa{
\lan r'|(x_1{-}x_{n+1})}\,.
\ee
We note that in computing the above Jacobian we should have included an absolute value e.g.~in $(x_1^2)^3$. We argue later in Section~\ref{sec:sign_functions} that we can assume $x_i^2>0$ for all $i$ (see \eqref{eq:allpos}).

\medskip
\noindent
{\it Inversion of
$\delta^{(4)}
(\lan r|(\theta_1{-}\theta_{n+1})^A)$.}
Under inversion
$\theta_i\to x_i^{-1}\theta_i$:
\be
\lan r|(\theta_1{-}\theta_{n+1})^A
\;\longrightarrow\;
[r|(x_1^{-1}\theta_1
- x_{n+1}^{-1}\theta_{n+1})^A\,.
\ee
Writing this as
$[r|\,x_1^{-1}(\theta_1{-}\theta_{n+1})^A
+ [r|(x_1^{-1}{-}x_{n+1}^{-1})
\theta_{n+1}^A$,
the second term is proportional to
$(x_1{-}x_{n+1})$,
which vanishes on the bosonic delta
support. Therefore
\be
[r|\,x_1^{-1}(\theta_1{-}\theta_{n+1})^A
= \frac{1}{x_1^2}\,
\lan r'|(\theta_1{-}\theta_{n+1})^A\, , 
\ee
and hence 
\be\label{eq:delta4_weight}
I\!\left[\delta^{(4)}\!\pa{
\lan r|(\theta_1{-}\theta_{n+1})^A}
\right]
= \frac{1}{(x_1^2)^4}\;
\delta^{(4)}\!\pa{
\lan r'|(\theta_1{-}\theta_{n+1})^A}\,.
\ee
\medskip
\noindent
{\it Inversion of $\vev{ri}$.}
From~\eqref{eq:inversions}, also transforming $\langle r| \to [r|$, we get 
\be\label{eq:I_ri}
I(\vev{ri})
= [r|_{\dot\alpha}(x_i^{-1})^{\dot\alpha\beta}
\lambda_{i,\beta}
= \frac{1}{x_i^2}\,[r|\,x_i\,|i\ran\,.
\ee
On the collinear support,
using
$x_i = x_1 - \sum_{a<i}p_a$
where $p_a = \lambda_a\lt_a$, 
\be\label{eq:rxi}
[r|\,x_i\,|i\ran
= [r|\,x_1\,|i\ran
- \sum_{a<i}[r\,\lt_a]\,
\underbrace{\vev{a\;i}}_{=0}
= [r|\,x_1\,|i\ran
\coloneqq \vev{r'i}\,,
\ee
since $\vev{ai}=0$ for all pairs
on the collinear support, and with $|r^\prime\rangle$   defined in  \eqref{eq:rhat2}. 
Therefore
\be\label{eq:I_ri_result}
I(\vev{ri})
= \frac{\vev{r'i}}{x_i^2}\,,
\ee
with $\vev{r'i} = [r|\,x_1\,|i\ran$.
As a consequence, 
the spinorial prefactor  in \eqref{eq:superamp} transforms as
\be
I\!\left[\frac{1}{|\vev{rn}\vev{r1}|}\right]
= \frac{x_1^2\,x_n^2}
       {|\vev{r'n}\vev{r'1}|}\,,
\ee
where as before we have assumed $x_i^2>0$. Finally, from~\eqref{eq:inversions}--\eqref{eq:bracket_inv},
we have 
$I[\delta(\vev{a\;a{+}1})]
= x_a^2\,\delta(\vev{a\;a{+}1})$,
contributing $\prod_{a=1}^{n-1}x_a^2$
in total.

\medskip
\noindent
{\it Full result.}
We collect below  the inversion weights
of all factors that do not involve
the stripped amplitude:
\begin{center}
\renewcommand{\arraystretch}{1.3}
\begin{tabular}{|l|c|}
\hline
\textbf{Factor} & \textbf{Weight under $I$} \\
\hline
$\displaystyle\frac{1}{|\vev{r1}\vev{rn}|}
\prod_{a=1}^{n-1}\delta(\vev{a\;a{+}1})$
& $(x_1^2)^2\,x_2^2\cdots x_{n-1}^2\,x_n^2$
\\[10pt]
$\delta^{(2)}(\lan r|(x_1{-}x_{n+1}))$
& $(x_1^2)^3$ \\[6pt]
$\delta^{(4)}(\lan r|
(\theta_1{-}\theta_{n+1})^A)$
& $(x_1^2)^{-4}$ \\[6pt]
\hline
\end{tabular}
\end{center}
 Hence,  for the 
covariance \eqref{eq:expected_weight} to hold we need to prove that 
\be\label{eq:dsc_final}
I[\tilde{A}_{1\ldots n}] = \tilde{A}_{1\ldots n}\,,
\ee
i.e.\ the stripped amplitude must have
unit weight under inversion.
This is established 
in the next section, 
where we prove~\eqref{eq:dsc_final}
at all~$n$ by rewriting the sign functions
in an arbitrary frame. 

\subsection{Invariance of the stripped amplitude
under dual conformal inversions}
\label{sec:sign_functions}

We now prove  the invariance \eqref{eq:dsc_final} of the stripped amplitude
$\tilde{A}_{1\ldots n}$
under  dual conformal inversion.  
In the general $n$ case, the proof combines the  transformation
 for spinor chains   
$\langle r|\,P_L\,P_R\,|r\rangle$
built from differences of  consecutive dual momenta 
 with the $|r'\ran$~trick introduced
in the previous section. 
It is instructive however to  first discuss  the three- and four-point cases before considering an arbitrary number of particles.

\medskip
\noindent
{\it Three-point superamplitude.} 
We know already from the work of \cite{Brandhuber:2008pf}
that the three-point $\overline{\rm MHV}$ superamplitude
is dual conformal covariant, but it is instructive
to confirm this when the superamplitude
is written in the form~\eqref{eq:3-pt-super-deltas}.
Because $\Delta^{(2)}$ is built with absolute values
and the bosonic delta Jacobian also produces
absolute values,
the inversion weight of all factors
other than~$\tilde{A}_{123}$
is $|x_1^2\,x_2^2\,x_3^2|$.
The stripped amplitude
$\tilde{A}_{123}
= \sg(\vev{r2}\vev{r3})\,\sg_{23}$ from \eqref{eq:Atilde123}
picks up
$\sg(x_2^2\,x_3^2)$ from
$I(\vev{ri}) = \vev{r'i}/x_i^2$
and $\sg(x_1^2)$ from
$I([23]) = [23]/x_1^2$,
giving
$I[\tilde{A}_{123}]
= \sg(x_1^2\,x_2^2\,x_3^2)\,
\tilde{A}_{123}$.
Combining,
\be
I[\cA_3^{(-)}]
= \sg(x_1^2\,x_2^2\,x_3^2)\,
|x_1^2\,x_2^2\,x_3^2|\;\cA_3^{(-)}
= x_1^2\,x_2^2\,x_3^2\;\cA_3^{(-)}\,,
\ee
confirming dual conformal covariance
with no assumption on the signs of the~$x_i^2$.

\medskip
\noindent
{\it Four-point superamplitude.}
At four points the situation is more subtle.
The stripped amplitude
$\tilde{A}_{1234}$ in \eqref{eq:4ptAtfinal}
is a sum of two terms.
Under inversion,
the first acquires $\sg(x_1^2\,x_2^2\,x_3^2\,x_4^2)\sg(x_1^2\,x_3^2)$
(from the pair $[12]$, $[34]$
via~\eqref{eq:bracket_inv}
and the $\vev{ri}$ factors),
while the second acquires $\sg(x_1^2\,x_2^2\,x_3^2\,x_4^2)\sg(x_2^2\,x_4^2)$.
Since the two are not the same, invariance of $\tilde{A}_{1\ldots 4}$ is only achieved  by making some assumptions on the signs of the $x_i^2$. In this respect, 
we note that allowing the $x_i^2$ to have independent signs would flip the sign of Mandelstam invariants, e.g.~$s_{12}=x_{13}^2\to  x_{13}^2 /(x_1^2 x_3^2) $, and if $x_1^2 x_3^2 <0$  the $i\varepsilon$ prescription of the  propagator $s_{12} + i \varepsilon$ would be changed. For a generic kinematic configuration  this is irrelevant, but for the 
special half-collinear kinematics the $i \varepsilon$ prescriptions have to be kept in place. 
We now argue that 
we can avoid this issue by choosing
the origin of dual momentum space appropriately such that all the $x_i^2$ are positive. In that case, $\sg(x_1^2\,x_3^2) = \sg(x_2^2\,x_4^2) = 1$ and $\tilde{A}_{1234}$ is manifestly invariant.

\medskip
\noindent
{\it Choice of region.}
We note that in the half-collinear kinematics we are considering, we can write the generic dual coordinate $x_i$ as 
$x_i = x_1 - \lambda\sum_{a<i}c_a\lt_a$. We then have 
$x_i^2 = x_1^2 - \mu_1\Cdot T_i$
where $\mu_1 = x_1\lambda$ and
$T_i = \sum_{a<i}c_a \lt_a$.
By rescaling $x_1\to s\,x_1$
with $|s|$ large, all $x_i^2$ can
be made to share the sign of $x_1^2$.
We choose the region where
\be\label{eq:allpos}
x_i^2 > 0 \qquad\text{for all }i=1,\ldots,n\,.
\ee
This is legitimate since the stripped amplitude
is independent of~$x_1$.

\medskip
\noindent
{\it Covariant form of the sign functions.}
We can now move on to  the $n$-point case. The stripped amplitude $\tilde{A}_{1\ldots n}$ is built out of   sign functions
$\sg(\bev{\tLam_L,\tLam_R})$
and step functions of the form 
$\Theta(\bev{\tLam_L,\tLam_R}
/\bev{\tLam_S,\tLam_T})$ (which can also  be rewritten as sign functions), where 
the quantities  $\tLam_A$ were introduced earlier in \eqref{eq:tLam}.  Note that in the frame \eqref{eq:frame} we have $\vev{ri} = -1$ and hence $\tLam_A = -\tilde{\lambda}_A$.
The blocks $L$ and  $R$ are also consecutive in the
colour ordering.
The fact that these blocks are made of consecutive momenta in colour ordering has important consequences from the point of view of dual conformal symmetry,  as we now show. 
For any  two consecutive
blocks $L=\{i,\ldots,j\}$
and $R=\{j{+}1,\ldots,k\}$, we can write 
\be\label{eq:chain_id}
\bev{\tLam_L,\tLam_R}
= \sum_{a\in L}\sum_{b\in R}
  \vev{ra}\vev{rb}\bev{ab}
= -\sum_{a\in L}\sum_{b\in R}
  \vev{ra}\bev{ab}\vev{br}
= -\langle r|\,P_L\,P_R\,|r\rangle\, , 
\ee
with $P_L\coloneq \sum_{a\in L} p_a$ and $P_R\coloneq \sum_{b\in R} p_b$. 
In dual coordinates $P_L = x_A-x_B$,
$P_R = x_B-x_C$, and every sign function
takes the form
\be\label{eq:fABC_def}
s_{L,R} \coloneq \sg(\bev{\tLam_L,\tLam_R})
= -\sg\!\big(\langle r|\,x_{AB}\,x_{BC}\,|r\rangle\big)\,,
\ee
where $x_{AB} \coloneq x_A - x_B$
and  where the two consecutive blocks share the
vertex~$x_B$.

\medskip
\noindent
{\it Dual inversions of spinor chains.}
Dual conformal inversions map
$x_i \to x_i^{-1}$ and hence
\be x_{AB}  \to
 -x_A^{-1}\,x_{AB}\,x_B^{-1}= - \frac{x_A x_{AB} x_B}{x_A^2 x_B^2}\, . \ee
For  a spinor chain
with particle spinors at the endpoints one would get \cite{Drummond:2008vq}
\be\label{eq:Drummond:2008vq_chain}
I\!\Big[\langle A|\,x_{AB}\,x_{BC}\,|C\rangle\Big]
= \frac{\langle A|\,x_{AB}\,x_{BC}\,|C\rangle}
       {x_A^2\;x_B^2\;x_C^2}\,.
\ee
In our case however the endpoints carry the
reference spinor $\langle r|$ and $|r\rangle$
rather than particle spinors, and we need to modify this argument. 
Under inversion $\langle r|\to [r|$
(see \eqref{eq:rhat1}),
and we then note that 
\be
\begin{split}
\langle r| x_{AB} \cdots &\to -\frac{1}{x_A^2 x_B^2}[ r | x_Ax_{AB}x_B  = -\frac{1}{x_A^2 x_B^2}[ r | (x_1  - x_{1A}) x_{AB}x_B
  = -\frac{1}{x_A^2 x_B^2}[ r | x_1  x_{AB}x_B
\\ &  = 
 - \frac{1}{x_A^2 x_B^2}\langle r^\prime |   x_{AB}x_B \, , 
\end{split}
\ee
where we have used the definition of $\langle r'|$ given in  \eqref{eq:rhat2}. Importantly, we also used that 
$x_{1A} x_{AB} = \sum_{i=1}^{A-1} \sum_{j=A}^{B-1} |i]\vev{ij}[j|=0$, since $\vev{ij}=0$ in the half-collinear kinematics. Similarly one can see that 
 \be
 x_{AB}|r\rangle \to - \frac{1}{x_A^2x_B^2 }x_Ax_{AB}|r^\prime\rangle\, .
 \ee
Hence, for a typical spinor chain appearing as argument in a sign function we have  
\be\label{eq:inv_chain}
I\!\Big[\langle r|\,x_{AB}\,x_{BC}\,|r\rangle\Big]
= \frac{\langle r'|\,x_{AB}\,x_{BC}\,|r'\rangle}
       {x_A^2\;x_B^2\;x_C^2}\,.
\ee

\medskip
\noindent
{\it Proof of invariance.}
By~\eqref{eq:inv_chain}, each
sign-function argument picks up a
positive weight in the
region~\eqref{eq:allpos}:
\be
\sg\!\big(I[f_{ABC}]\big)
= \sg\!\bigg(\frac{f_{ABC}(r')}{x_A^2 x_B^2 x_C^2}\bigg)
= \sg\!\big(f_{ABC}(r')\big)\,,
\qquad
f_{ABC} \coloneq \langle r|\,x_{AB}\,x_{BC}\,|r\rangle\,.
\ee
Thus we have found that 
\be
I\big[\tilde{A}_{1\ldots n}(r)\big] = \tilde{A}_{1\ldots n}(r')\,.
\ee
It is immediate to see that 
$\tilde{A}_{1\ldots n}(r') = \tilde{A}_{1\ldots n}(r)$ by rewriting $\vev{ri}= c_i \vev{r\lambda}$ and noting that  e.g. $\sg (\bev{\tLam_L,\tLam_R})
= \sg \Big( \vev{r\lambda}^2\sum_{a\in L}\sum_{b\in R}
  c_a c_b\bev{ab}\Big) = \sg \Big( \sum_{a\in L}\sum_{b\in R}
  c_a c_b\bev{ab}\Big)$, 
and hence 
\be\label{eq:A_invariance}
I[\tilde{A}_{1\ldots n}] = \tilde{A}_{1\ldots n}
\qquad\text{for all }n\,.
\ee
Finally, combining the results of this and the previous sections, we have shown that  
\be\label{eq:dci_final_finite}
I\!\left[\cA_n^{(-)}\right]
= \prod_{k=1}^n x_k^2\;\cA_n^{(-)}\,.
\ee

\medskip
\noindent
{\it Dual superconformal symmetry.}
The remaining dual superconformal generators
-- the dual Poincar\'e supersymmetries
$\bar{Q}$ and the dual special
superconformal generators $\bar{S}$ --
when restricted to  on-shell superspace,
coincide with the ordinary superconformal
generators $\bsb$ and $\bq$
respectively~\cite{Drummond:2008vq},
which are symmetries of any tree-level
$\cN=4$ super Yang-Mills amplitude.
Therefore the full dual superconformal symmetry
is established for arbitrary $n$.

\section{Grassmannian interpretation}
\label{sec:grassmannian}

In the Grassmannian formulation of $\cN=4$
SYM~\cite{Arkani-Hamed:2009ljj,Arkani-Hamed:2012zlh},
tree-level N${}^{k-2}$MHV superamplitudes
are obtained from an integral over
$\mathrm{Gr}(k,n)$.
The single-minus superamplitude has
Grassmann degree~4, corresponding
to $k=1$, for which
$\mathrm{Gr}(1,n) = \mathbb{CP}^{n-1}$
with the $C$-matrix being a single row
$(c_1,\ldots,c_n)$
modulo $\mathrm{GL}(1)$.
The $\mathrm{Gr}(1,n)$ integral is
\be\label{eq:Gr1n}
F_n \coloneq \int
\frac{d^nc\,/\,\mathrm{GL}(1)}
{c_1\,c_2\cdots c_n}\;
\delta^{(2)}\!\pa{\sum_i c_i\lt_i}\;
\delta^{2(n-1)}\!\pa{C^\perp\!\cdot\lambda}\;
\delta^{(4)}\!\pa{\sum_i c_i\eta_i^A}\,,
\ee
where $C^\perp$ is an $(n{-}1)\times n$
matrix whose rows span
the orthogonal complement of~$C$.
The constraint $C^\perp\!\cdot\lambda=0$
forces $\lambda_i = c_i\lambda$
for a common spinor~$\lambda$,
i.e.\ the half-collinear
condition~\eqref{eq:halfcoll}.

We  can now evaluate $F_n$ in split signature
using real delta functions.
Fix $\mathrm{GL}(1)$ by setting $c_1=1$
and take
\be\label{eq:Cperp}
C^\perp = \begin{pmatrix}
-c_2 & 1 & 0 & \cdots & 0\\
-c_3 & 0 & 1 & \cdots & 0\\
\vdots & & & \ddots \\
-c_n & 0 & 0 & \cdots & 1
\end{pmatrix}\,,
\ee
so $C^\perp\!\cdot\lambda=0$ imposes
$\lambda_a^\alpha - c_a\lambda_1^\alpha = 0$
for each $a=2,\ldots,n$ and $\alpha=1,2$.
For each~$a$, one component
(say $\alpha_0$ with
$\lambda_1^{\alpha_0}\neq 0$)
fixes $c_a = \lambda_a^{\alpha_0}/\lambda_1^{\alpha_0}$,
producing a Jacobian $1/|\lambda_1^{\alpha_0}|$.
The remaining component becomes
$|\lambda_1^{\alpha_0}|\;\delta(\vev{1a})$.
The two factors cancel pairwise,
giving
\be\label{eq:grass_coll_deltas}
\int\prod_{a=2}^n dc_a\,
\delta^{2(n-1)}(C^\perp\!\cdot\lambda)
= \prod_{a=2}^n\delta(\vev{1a})\,,
\ee
with no residual Jacobian.
Writing the localised values covariantly,
$c_a^* = \vev{ra}/\vev{r1}$,
the denominator, momentum delta
and fermionic delta functions evaluate to
\be\label{eq:grass_denom}
\frac{1}{\prod_{i=1}^n c_i^*}
= \frac{\vev{r1}^{n-1}}
{\prod_{a=2}^n\vev{ra}}\,,
\ee
\be\label{eq:grass_mom}
\delta^{(2)}\!\pa{\sum_i c_i^*\lt_i}
= \vev{r1}^2\;\delta^{(2)}\!\pa{
\sum_i\vev{ri}\lt_{i\dot{\alpha}}}\,,
\ee
\be\label{eq:grass_ferm}
\delta^{(4)}\!\pa{\sum_i c_i^*\eta_i^A}
= \frac{1}{\vev{r1}^4}\;\delta^{(4)}\!\pa{
\sum_i\vev{ri}\eta_i^A}\,,
\ee
where the factor $\vev{r1}^{-4}$ arises because
the fermionic delta function is a degree-4
Grassmann polynomial linear in the~$c_i^*$.
Collecting powers of $\vev{r1}$:
$(n{-}1)+2-4 = n-3$, giving
\be\label{eq:Gr_localised}
F_n =
\frac{\vev{r1}^{n-3}}
{\prod_{a=2}^n\vev{ra}}\;
\prod_{a=2}^n\delta(\vev{1a})\;
\delta^{(2)}\!\pa{\sum_i\vev{ri}\lt_{i\dot{\alpha}}}\;
\delta^{(4)}\!\pa{\sum_i\vev{ri}\eta_i^A}
\,.
\ee
Comparing with the covariant
superamplitude~\eqref{eq:superamp}
at $k=1$:
\be\label{eq:Gr_comparison}
\cA_n^{(-)}
= i^{2-n}\,\tilde{A}_{1\ldots n}\;F_n\,.
\ee
The Grassmannian integral contains
no sign functions;
the stripped amplitude $\tilde{A}_{1\ldots n}$
is the entire discrepancy between
$F_n$ and the superamplitude.

A natural interpretation is that the Grassmannian description requires a more refined specification of the {\it real} integration contour. In split signature, where integrals over real cycles often lead to piecewise-constant quantities~\cite{Arkani-Hamed:2009hub}, one might expect such a refinement to involve a decomposition into chambers. These  are regions where the stripped amplitude $A_{1\ldots n}$ has a constant  value, and they are separated by walls, across which the value of  $A_{1\ldots n}$ jumps. 
This interpretation is also suggested by comparing the split-signature and complex localisations. In complex kinematics the Grassmannian is evaluated using holomorphic delta functions~\cite{Witten:2003nn,Roiban:2004yf}, which produce algebraic Jacobians $J^{-1}$ with no absolute values. At each step of the Berends--Giele recursion of~\cite{Guevara:2026qzd}, the ratio $J/|J|$ that produces sign functions in split signature is replaced by $J/J=1$, so every sign function reduces to a fixed $\pm1$ and the stripped amplitude takes a constant value in $\{-1,0,+1\}$ throughout the collinear locus. In split signature, by contrast, the passage to real delta functions replaces $J^{-1}$ by $|J|^{-1}$, and the sign functions $\mathrm{sg}(\bev{ij})$ measure precisely the discrepancy. The chamber structure of the stripped amplitude is the imprint of this replacement. We do not attempt to develop such a contour prescription here. Nevertheless, the chamber structure of $\tilde{A}_{1\ldots n}$ appears suggestive: the regions on which the stripped amplitude is constant are precisely the connected components of $\mathbb{RP}^{n-1}$.

Despite the absence of sign functions
in the direct localisation,
the Grassmannian provides a natural
geometric framework for the
chamber structure of $A_{1\ldots n}$.
In the amplituhedron
programme~\cite{Arkani-Hamed:2013jha},
amplitudes are canonical forms
on positive geometries
defined by the image of
$\mathrm{Gr}_+(k,n)$
under the kinematic map.
For $k=1$,
the positive Grassmannian
$\mathrm{Gr}_+(1,n)$
is the subset of~$\mathbb{RP}^{n-1}$
with all $c_i>0$ (or all $c_i<0$),
and its image under the collinear
kinematic data is a single cell.
The boundaries of this cell
are the walls $c_i=0$,
which on the collinear support
correspond to $\vev{ri}=0$ for some~$i$.
More generally, the chambers
of the stripped amplitude, that is 
the connected components of
$\mathbb{RP}^{n-1}$ minus
the coordinate hyperplanes $c_i=0$
and the composite walls
$\sum_{a\in I}c_a = 0$, 
are exactly the regions
in which $A_{1\ldots n}$
takes a fixed integer value.
The sign functions $\sg_{ij}$
track the relative signs
of pairs $c_i$, $c_j$,
i.e.\ the position within
$\mathbb{RP}^{n-1}$
relative to the coordinate
hyperplanes.
In this sense the chamber structure
of the single-minus amplitude
is the positive geometry
of the degenerate amplituhedron
at $k=1$.

\section{Single-minus amplitudes in \texorpdfstring{$\cN=8$}{N=8} supergravity}
\label{sec:sugra}

Finally, we turn to single-minus graviton tree amplitudes and their supersymmetric completion in $\mathcal{N}=8$ supergravity.  
As argued in \cite{Guevara:2026qwa}, the single-minus amplitudes in pure gravity are non-vanishing in the half-collinear kinematics. Following precisely the same logic as in the $\mathcal{N}=4$ superamplitude case, we find the supersymmetric completion of the single-minus amplitude in $\mathcal{N}=8$ supergravity takes the following form, 
\be \label{eq:M-GR}
\mathcal{M}_n^{(-)}= i^{2-n} \frac{ \Delta^{(n-1)} (\{\lambda_i\}) } { \prod_{a=1}^{n} \vev{r a}^{2}}  \tilde{M}_{1\cdots n}  \, \delta^{(2)}  \Big(\sum_{i=1}^n \vev{ri}\lt_i \Big)   \delta^{(8)} \Big( \sum_{i=1}^n \vev{ri}\eta_i\Big) \, ,  
\ee
where the permutation-symmetric quantity  $\Delta^{(n-1)} (\{\lambda_i\})$ is defined in \eqref{eq:Delta}, and now $A=1, \ldots, 8$. 

We now comment on general properties of the stripped amplitude $\tilde{M}_{1\cdots n}$.  First, it is evident that the denominator and $\Delta^{(n-1)}(\{\lambda_i\})$ in \eqref{eq:M-GR} together give correct little-group weights for all the external particles (including signs), therefore  $\tilde{M}_{1\cdots n}$ should  not have any little-group weight in the external particles\footnote{This is in contrast with the stripped amplitude $M_{1\cdots n}$ introduced in  \cite{Guevara:2026qwa}, which  does not strictly have zero helicity weight on external particles due to sign functions.}.  On the other hand, it does carry weight $2(n{-}2)$ in $|r \rangle$, such that the superamplitude is $|r \rangle$ independent. 
Moreover, unlike $\tilde{A}_{1\cdots n}$ in gauge theory, $\tilde{M}_{1\cdots n}$ is dimensionful, scaling as $p^{3(n-2)}$; and rather than cyclic invariance, $\tilde{M}_{1\cdots n}$ has full permutation symmetry.%
\footnote{The fact that  $\Delta^{(n-1)}(\{\lambda_i\})$ is permutation symmetric was shown in Section~\ref{sec:definition}.}
These properties ensure that the amplitude $\mathcal{M}^{(-)}_n$ 
satisfies  the requirements of an $\cN=8$ supergravity amplitude.

 More precisely,  the stripped amplitude is constructed from the helicity-blind building blocks  $[\tLam_L, \tLam_R]$  introduced in \eqref{eq:rep1}; however, whereas $\tilde{A}_{1\ldots n}$ involves sign functions of these brackets, $\tilde{M}_{1\cdots n}$ involves their absolute values.  For $n=3$ we have 
\be \label{eq:M123}
\tilde{M}_{12 3}  = |\vev{r |1 \, 2| r } |  \, ,  
\ee
 and as a comparison, recall that $\tilde{A}_{12 3}  = -\sg (\vev{r |1\,  2| r } )$, as given in \eqref{eq:3ptAtfinal}. Following the same calculation as in Section~\ref{sec:3pt}, it is straightforward to verify that $\mathcal{M}_3^{(-)}$ is in agreement with the known expression for the three-point $\overline{\rm MHV}$ superamplitude in $\mathcal{N}=8$ supergravity
\be
\mathcal{M}_{\overline{\rm MHV}} = -i\, \frac{\delta^{(4)} (\sum_{i=1}^3 \lambda_i\lt_i)\delta^{(8)} ( \eta_1 [23] + \eta_2[31] + \eta_3[12])}{[12]^2 [23]^2 [31]^2} \, . 
\ee
For $n=4$ we have   
\be \label{eq:M1234}
\tilde{M}_{1\ldots 4}  = \frac{1}{2}\Big[ |\langle r| 1\,2|r \rangle  \langle r| 3\,4|r \rangle | + | \langle r| 1\,3|r \rangle  \langle r| 2\,4|r  \rangle | + 
| \langle r| 1\,4|r  \rangle  \langle r| 2\,3|r \rangle |
\Big]  \, ,
\ee
where the sum runs over the three pairings of $\{1,2,3,4\}$ into two pairs. This should be compared  with $\tilde{A}_{1234}  = 
\frac{1}{2}\Big[ \sg (\langle r| 12|r \rangle  \langle r| 34|r \rangle) + 
\sg( \langle r| 23|r \rangle  \langle r| 41|r  \rangle)
\Big]
$ from  \eqref{eq:4ptAtfinal}, which involves only two of the three pairings, consistent with cyclic rather than full permutation symmetry.    

In fact, it is straightforward to see that $\tilde{M}_{123}$ and $\tilde{M}_{1\ldots 4}$ are completely fixed (up to an overall factor) by the properties of $\tilde{M}_{1\cdots n}$ discussed above. Higher-point $\tilde{M}_{1\cdots n}$ may be obtained directly from the stripped amplitude $M_{1\cdots n}$ introduced in \cite{Guevara:2026qwa} (via a relation analogous to that between $\tilde{A}_{1\cdots n}$ and $A_{1\cdots n}$ in $\cN{=}4$ SYM). Once again, one may consider a specific frame and kinematic regime in which the expression of the stripped amplitude simplifies. Notably, the stripped amplitude admits a representation as a sum over Cayley trees, in direct analogy with the MHV case \cite{Bern:1998sv, Nguyen:2009jk, Hodges:2011wm, Feng:2012sy} (see also \cite{Bedford:2005yy,Cachazo:2005ca} for the computation of MHV gravity amplitudes using BCFW recursion).  How much of this structure is in fact fixed by physical properties of $\tilde{M}_{1\cdots n}$, as was in the case of  $n=3$ and $n=4$, remains to be understood.
We also note that the replacement of sign functions by absolute values in passing from $\tilde{A}_{1\ldots n}$ to $\tilde{M}_{1\ldots n}$ is suggestive of a double-copy structure, though it remains unclear what the two copies are and how they should be combined, particularly given the role of the prefactors.

\section{Summary and discussion}
\label{sec:discussion}

In this paper we constructed the $\cN{=}4$ single-minus superamplitude~\eqref{eq:superamp} in  $(2,2)$ signature. The superamplitude factorises into a collinear measure $\Delta^{(n-1)}$, a helicity-blind stripped amplitude $\tilde{A}_{1\ldots n}$, and supersymmetrised momentum conservation delta functions. We proved dual superconformal covariance of the  superamplitude at all multiplicities by rewriting the sign function arguments as spinor chains built from consecutive dual coordinate differences. We also analysed the $\mathrm{Gr}(1,n)$ Grassmannian integral, showing that the superamplitude factorises as $\cA_n^{(-)} = i^{2-n}\,\tilde{A}_{1\ldots n}\,F_n$ with $F_n$ free of sign functions. Finally, we wrote the $\cN{=}8$ supergravity single-minus superamplitude, which shares the same collinear measure $\Delta^{(n-1)}$ and supersymmetrised delta functions, but with the stripped amplitude $\tilde{M}_{1\ldots n}$ built from absolute values rather than sign functions of  helicity-blind brackets, and enjoying permutation symmetry.

 We close with a few  remarks on the broader context of these results and the questions they raise.
 
{\bf 1.} One lesson of our work is that standard on-shell constructions require a refinement in the distributional sector.  As we have discussed,  the amplitudes studied here follow directly from the Feynman rules with the standard $i\varepsilon$ prescription, however they are not captured by the usual BCFW construction and are not reproduced in full by the $\mathrm{Gr}(1,n)$ Grassmannian integral -- 
indeed we have  shown that it correctly reproduces the universal factor $F_n$ but not the piecewise-constant stripped amplitude $\tilde A_{1\ldots n}$. 
  An important open question is therefore to determine the complex counterpart of the stripped amplitude at arbitrary $n$. 
As discussed in Section~\ref{sec:grassmannian}, the sign functions in $\tilde A_{1\ldots n}$ are absent from the Grassmannian integral and arise only upon passing to real delta functions in split signature. This suggests that a more refined treatment of real contours may be required, with the chamber structure discussed there providing a natural starting point.  It would also be important   to identify the canonical form on $\mathrm{Gr}_+(1,n)$ that reproduces the stripped amplitude in each chamber and to understand how it arises as a degeneration of the amplituhedron~\cite{Arkani-Hamed:2013jha}.

{\bf 2.} The amplitudes constructed here are supported on a locus of measure zero and do
not themselves contribute to cross sections.  There are, however, several
settings in which distributional split-signature amplitudes may be expected to
play a physical role, and identifying such a role seems to us the natural
continuation of this work: 
\begin{itemize}
    \item[{\bf a.}]
The first is self-dual Yang--Mills theory.  The
single-minus configuration is the first one lying outside the self-dual sector,
which is classically integrable and whose quantum structure is famously
subtle: the one-loop all-plus and single-minus amplitudes are nonvanishing
rational functions, and Bardeen's conjecture interprets the one-loop all-plus
sector as an anomaly of the self-dual
theory~\cite{Mahlon:1993si,Bardeen:1995gk,Cangemi:1996rx,Chalmers:1996rq}.
Understanding how the distributional tree amplitudes studied here fit into that
picture, and in particular whether they leave an imprint on the known rational
one-loop amplitudes, is a natural first question.

 \item[{\bf b.}]
A second setting is twistor theory and the twistor
string~\cite{Witten:2003nn,Roiban:2004yf}, which are formulated in split
signature precisely because the reality conditions available there turn the 
twistor transform into a straightforward Fourier transform. In the twistor
classification of~\cite{Witten:2003nn} the single-minus amplitudes have support on  
degree-zero curves, in which the holomorphic curve degenerates to a point,
and  it would be interesting  to develop the twistor-space description of the
superamplitudes constructed here directly.  

\item[{\bf c.}]
A third is celestial
holography~\cite{Pasterski:2021rjz,Atanasov:2021oyu}, where scattering is again
naturally formulated in split signature, and where distributional,
collinear-supported data are the rule rather than the exception.

\end{itemize}

{\bf 3.}
Establishing full Yangian invariance
in the sense of~\cite{Drummond:2009fd}
would require verifying
explicitly the remaining level-zero generators
(conformal boost and special superconformal)
beyond what we have checked here;
we expect this to follow from the superconformal algebra,
given that the $n{=}3$ superamplitude coincides with
the $\overline{\rm MHV}$ superamplitude
of~\cite{Brandhuber:2008pf} for which Yangian invariance
is understood.

{\bf 4.} Finally, the gravitational side of this programme deserves further attention.
The $\cN{=}8$ superamplitude of Section~\ref{sec:sugra} is built from the same
permutation-invariant measure $\Delta^{(n-1)}$ and the same supersymmetrised
delta functions as its gauge-theory counterpart, with sign functions replaced by
absolute values of the same helicity-blind brackets.  As noted there, this
replacement is suggestive of a double-copy structure whose two copies
remain unidentified.  Making that structure precise, and obtaining the
higher-point stripped amplitudes $\tilde{M}_{1\ldots n}$ in closed form, would be
a natural next step, as would extending the analysis of the graviton amplitudes
of~\cite{Guevara:2026qwa} along the lines followed here.

We hope to return to some of these questions in future work.

\section*{Acknowledgements}

We thank Costis Papageorgakis and David Vegh
for  several useful discussions. 
This work was supported by the Science and Technology Facilities Council (STFC) Consolidated Grant ST/X00063X/1 \textit{``Amplitudes, Strings  \& Duality''}. CW is supported by a Royal Society University Research Fellowship No.~URF$\backslash$R$\backslash$221015.   No new data were generated or analysed during this study.


\bibliographystyle{JHEP}
\bibliography{refs}

\end{document}